\documentclass[12pt]{article}%
\usepackage{amssymb}
\usepackage{amsmath}
\usepackage{amstext}
\usepackage{natbib}
\usepackage{fancyhdr}
\usepackage[hang,flushmargin,multiple]{footmisc}
\usepackage{geometry}
\usepackage{mathpazo}
\usepackage[doublespacing]{setspace}
\usepackage[multiple]{footmisc}
\usepackage{amsfonts}
\usepackage{graphicx}%
\usepackage{lscape}
\usepackage{booktabs}

\newcounter{def}

\newcounter{assumption}

\usepackage[colorlinks]{hyperref}
\hypersetup{colorlinks = {true},
urlcolor={black},
citecolor={black},
linkcolor={black},
anchorcolor={black}
filecolor={black}
}
\bibpunct{\color{black}{(}}{\color{black}{)}}{;}{a}{}{,}
\usepackage{color}
\usepackage[compact]{titlesec}
\titlespacing{\section}{0pt}{10pt}{10pt}
\titlespacing{\subsection}{0pt}{9pt}{8pt}
\titlespacing{\subsubsection}{0pt}{9pt}{6pt}
\titleformat{\section}
{\normalfont\Large\bfseries}
{\thesection.}{0.5em}{}
\titleformat{\subsection}
{\normalfont\large\bfseries}
{\thesubsection}{0.5em}{}
\titleformat{\subsubsection}
{\normalfont\normalsize\itshape}
{\thesubsubsection.}{0.5em}{}
\usepackage{comment}
\makeatletter
\g@addto@macro\normalsize{%
  \setlength\abovedisplayskip{6pt}
  \setlength\belowdisplayskip{6pt}
  \setlength\abovedisplayshortskip{6pt}
  \setlength\belowdisplayshortskip{6pt}
}
\makeatother

\begin{document} 

\singlespacing
\thispagestyle{empty} \renewcommand{\footnoterule}{\rule{0pt}{0pt}} %
\title{Integrating Structural and Reduced-Form Methods  in Empirical Finance \bigskip \bigskip }

\author{Toni M. Whited \bigskip \bigskip \thanks{Toni M. Whited is from the University of Michigan and NBER. twhited@umich.edu. I thank Yuan Shi for her excellent research assistance.}}

\date{\today}
 
\maketitle

\textbf{Abstract:}  I discuss various ways in which inference based on the estimation of the parameters of statistical models (reduced-form estimation) can be combined with inference based on the estimation of the parameters of economic models (structural estimation). I discuss five basic categories of integration:  directly combining the two methods, using statistical models to simplify structural estimation, using structural estimation to extend the validity of reduced-form results, using reduced-form techniques to assess the external validity of structural estimations, and using structural estimation as a sample selection remedy. I illustrate each of these methods with examples from corporate finance, banking, and personal finance.  
 
\newpage
\setcounter{footnote}{0}
\renewcommand{\thefootnote}{\arabic{footnote}}

\setcounter{page}{1}

\doublespacing
\baselineskip=26pt
\renewcommand{\footnoterule}{\noindent \rule{2.23in}{0.62pt}}

\section{Introduction}

How can reduced-form and structural econometric methods complement each other to push knowledge forward?  The goal of this paper is to illustrate the synergies between these two seemingly disparate methodologies in the context of corporate finance, as well as the related fields of financial intermediation, personal finance, and real estate. This discussion rests on the back of long-standing methodological debates, dating back at least to \cite{Koopmans1947}, about the most effective ways to learn from data. This debate been hashed out several times in both economics and finance \citep{AngristPischke2010,Keane2010,Welch2013,StrebulaevWhited2013}, and both \cite{Heckman2010} and \cite{ToddWolpin2022} offer discussions of how to integrate structural modeling into empirical program evaluation. Given this background, it seems that the integration of reduced-form with structural work has been well examined. However, much of this discussion in economics has centered around program evaluation or randomized controlled trials (RCTs), while the area of finance has a paucity of programs or RCTs to evaluate.  Thus, the ways in which researchers have integrated different methodologies are both different from the methods used in economics and quite varied. 

Before providing a taxonomy of the methods for integrating reduced-form and structural methods, it is useful to start with working definitions, as the terms structural and reduced-form have several different meanings in different contexts, none of which are either right or wrong. The differences are purely semantic. For example, the theoretical credit risk models based on the framework in \cite{Merton1974} are often termed structural. Similarly, theoretical dynamic single-agent optimization models or dynamic equilibrium models of the economy are also often termed structural models. 

For the purposes of the discussion that follows, we use the term structural to describe empirical exercises that are guided by formal economic theory. Specifically, the objects to be estimated are the parameters of a formal economic model. As such, this type of empirical exercise typically starts with an economic model defined by assumptions about agents' preferences and constraints, firms' or institutions' technologies, an objective to be optimized, and often a notion of equilibrium. The model then generates predictions about the relationships between observable, often endogenous variables.  Structural estimation then uses these relationships to ascertain whether optimal decisions implied by a model resemble actual decisions by agents. These agents can include firms, banks, households, venture capitalists, or government agencies. The outputs of this type of exercise are an assessment of the fit of the model and estimates of the model's parameters, which can include parameters that describe preferences, technologies, or frictions.  Examples of the types of questions that can be answered include the effect of equity market misvaluation on corporate policies \citep{WarusawitharanaWhited2016} or the effect of risk on entrepreneurial activity \citep{Catherine2022}.

 The term ``reduced-form'' similarly has more than one meaning. Formally, it refers to a statistical model that has been derived formally from a theoretical model in such a way that endogenous variables are a function of an unobserved error term and a set of exogenous observable variables.  A more common definition is a statistical model that describes the relationships between two or more random variables; for example, a linear regression. Of course, nonlinear models such as logit or negative binomial regressions also fall into this category.  This framework can answer a large variety of questions. How do shocks to banks affect firm investment \citep{AmitiWeinstein2018}? How do investor protection laws affect corporate decisions \citep{Agrawal2013}? Does the securitization of loans affect banks' incentives to screen borrowers \citep{KeysMukherjeeSeruVig2010}? How has executive compensation evolved in the last century \citep{FrydmanSaks2006}?  All of these studies are descriptive in the sense that they estimate relationships between economic variables, and in some cases, studies such as these can claim a causal relationship, typically when either formal economic theory or informal economic intuition guides an assumption about the error term in a regression.

Interestingly, the goal of both methods is to uncover causal effects between two economic variables or outcomes.  Because a causal effect can be defined as a counterfactual statement in which one variable affects another, artificially holding everything else constant, the most obvious example is a well-specified OLS regression.  As long as the error term is credibly uncorrelated with the regressors, then the slope coefficient on a regressor of interest represents a causal effect, holding all other variables constant.  Quasi-experimental designs, such as difference-in-differences setups, use similar and often much more economically plausible assumptions to offer statistical results with a causal interpretation. 

The direct estimation of economic models also can allow for causal interpretations, but the path is typically different. For example, after the estimation of a model's parameters, a researcher can introduce a counterfactual shock or change a model parameter counterfactually to understand the effect of a shock or a parameter on an economic outcome.  

The goals of these two types of empirical methods are clearly related, yet they differ in one important dimension: the types of variables involved in causal statements. While reduced-form exercises usually study causal relations between observable variables, structural studies often examine the effects of unobservables by imposing sufficient structure on the econometric model. The concurrence of goals and the variety in the types of outcomes studied opens paths for combining these two broad methods. 

Five types of integration have been featured in corporate finance.  First, a model can incorporate elasticities estimated from a reduced-form regression. Second, part of the model is simplified via a reduced-form regression to reduce complexity. Third, a model can extend the external validity of a reduced-form result. Fourth, a reduced-form regression can serve as a check of external validity or, similarly, as motivation for constructing a model. Fifth, models can be used to address sample selection problems in regressions. 
The rest of this paper discusses these five types of integration using illustrative examples. 

\section{Background} 

\subsection{Notation}

For the purpose of exposition, it is useful to write down a simple framework that can be used to illustrate these types of integration. In general terms, for data that consist of observations indexed both by cross-sectional units, $i$, and time, $t$, one can think of a reduced-form statistical model as a relationship between a scalar outcome , $y_{it}$, a vector of other variables, $x_{it}$, a vector of parameters, $\beta$, and a scalar error term, $u_{it}$: \begin{equation}\label{eq:rf}
y_{it} = F(x_{it},\beta,u_{it}). 
\end{equation}
Equation \eqref{eq:rf} encompasses many different commonly used reduced-form statistical techniques. For example, if $y_{it}$ is a scalar, the function $F(\cdot)$ is linear, and $u_{it}$ is uncorrelated with all of the elements of $x_{it}$, then \eqref{eq:rf} is a well-specified ordinary least squares regression. Equation \eqref{eq:rf} can encompass a wide variety of statistical models by appropriate definitions of $F(\cdot)$ and appropriate assumptions about the correlation structure of $(y_{it},x_{it},u_{it})$.  For example, instrumental variables models, difference-in-differences models, logistic regressions, probit models, and negative binomial models all can be characterized by \eqref{eq:rf}. 

Structural estimation starts with an economic model, the solution to which is often a decision rule that takes the form:
\begin{equation}\label{eq:st}
y_{it} = G(y_{-it},x_{it},b,z_{it}).
\end{equation} 
In equation \eqref{eq:st}, $y_{it}$ is the optimal decision of agent $i$ at time $t$, $y_{-it}$ are the optimal decisions of other, possibly competing agents, $(x_{it}, z_{it})$ are vectors of variables describing the state that agent $i$ faces at time $t$. While $x_{it}$ can be influenced by past decisions, $y_{is}, s<t$, the vector $z_{it}$ cannot be influenced by any agents' actions. The vector $b$ contains economic parameters such as risk-aversion or firm technological parameters.  This vector typically does not contain explicit elasticities between $y_{it}$ and any of the elements of $(x_{it}, z_{it})$.  Equation \eqref{eq:st} is often the solution to an economic model that consists of pages of equations describing the economic environment and the objective function that agents maximize. 

One common misconception of this type of economic model is that the point is to infer causality between $z_{it}$ and $y_{it}$, and that this type of exercise is trivial because $z_{it}$ is assumed to be exogenous to agents' decisions. Of course, many counterfactual exercises revolve around setting elements of $b$ to zero. Even when the goal of a counterfactual is to examine responses such as $\partial y_{it} /\partial z_{it}$, the endogenous and often data-informed form of $G(\cdot)$ modulates the intensity of this response. 

The goal of structural estimation is to obtain an estimate of the parameter vector $b$. While many different methods have been used in a wide variety of studies, a few have been dominant. Occasionally, it is possible to derive a simple reduced-form statistical model from \eqref{eq:st} and use conventional statistical techniques to recover the parameter vector, $b$.  A classic example is the estimation of a consumption Euler equation using nonlinear instrumental variables as in \cite{HansenSingleton1982}, in which $b$ is the parameter in a constant relative risk aversion utility function. Another example is in \cite{DavisFisherWhited2014}, in which \eqref{eq:st} is a dynamic equilibrium model of an economy with agglomeration externalities, \eqref{eq:rf} is an instrumental variables regression of $y_{it}$ on $x_{it}$, and the element of $b$ that quantifies the externality is a function of $\beta$ and other easily estimated elements of $b$.  

In other instances, it is possible to derive a likelihood function from the economic model \eqref{eq:st}, although the likelihood function need not have a closed-form solution, as in the seminal work of \cite{Rust1987} or, more recently, in the work on financial frictions in \cite{KaraivanovTownsend2014}. In these cases, a numerical solution for the likelihood function is maximized using a nested fixed-point algorithm. 

The most popular methods in corporate finance for estimating the parameter vector $b$ are the analogy methods that fall under the umbrella category of indirect inference.  This general estimation principle uses an auxiliary statistical model of the form \eqref{eq:rf} that can be estimated using both actual data and data that are simulated from an economic model of the form \eqref{eq:st}. The econometric criterion function to be minimized typically measures the distance between the auxiliary parameter vector estimated with real data and the same parameter vector estimated with model-simulated data. The parameter vector from the economic model, $b$, can be inferred from the statistical model parameter, $\beta$, because of an assumed unique mapping from $b$ to a subvector of $\beta$ of the same dimension. While the auxiliary statistical model can take the form of a likelihood function \citep{GallantTauchen1996}, nearly all of the applications of this method to corporate finance fall under the category of simulated method of moments (SMM), which uses a list of moments as an auxiliary model or simulated minimum distance (SMD), which uses a list of functions of moments \citep[e.g.,][]{HennessyWhited2005,Taylor2010,BazdreschKahnWhited2018}. 

\subsection{Identification} 

Because the topic of identification arises in several of the examples that follow and because this term has different meanings in different contexts, it is worth discussing working definitions. In reduced-form studies, the goal is to estimate slope coefficients. While discussion of the identification of these slope coefficients has a long history dating back to the founding of the Cowles Foundation \citep{Christ1994}, and while identification in this sense typically rests on assumptions about the correlation between $u_{it}$ and $x_{it}$ in \eqref{eq:rf}, in modern parlance, the term often refers to something else. Specifically, it refers to assumptions about the economic environment that permit the association of a direction with a slope coefficient, as in the statement, ``$x_{it}$ causes $y_{it}$.'' While exogenous variation in one or more elements of $x_{it}$ is useful for obtaining a directional interpretation of an estimated regression coefficient, $\beta$, it is less useful for understanding the deeper behavioral parameters, $b$, that in turn drive $\beta$.  

In contrast, structural work does not usually identify a direction to be associated with an elasticity. Instead, the goal is the estimation of parameters describing economic agents, and the notion of identification is closer to the traditional econometric definition of an econometric loss function with a unique minimum, that is, no flat spots. In the context of SMM or SMD, this condition translates directly into a unique mapping from parameters to the moments or functions of moments used in the estimation.

\section{Elasticities and Models}

The first type of integration is based on using a regression elasticity to help identify a parameter in an SMD estimation. This type of exercise typically starts with an empirical setting based on a clean natural experiment or exogenous shock to an outcome of interest. The statistical model \eqref{eq:rf} is then a regression that incorporates a shock, uses an exogenous shock as an instrument, or is a comparison between treated and control groups in a natural experiment. An economic model as in \eqref{eq:st} is then built to feature this exogenous shock or experimental setting, and the parameters of the economic model, $b$, are estimated. 

Why might this type of research design be interesting or informative?  First, regression-based estimates of elasticities, $\beta$ in \eqref{eq:rf}, are often functions of more than one parameter, $b$ in \eqref{eq:st}, that describes the behavior of economic agents. Thus, modeling this behavior and estimating the model parameters, $b$, allows a researcher to understand the features of the economic environment that help determine the magnitude or sign of the elasticity, $\beta$. Conversely, although an element of $\beta$ is rarely a function of only one element of $b$, using the statistical model as a target to match in an SMD estimation can offer useful identifying information.  

Second, the model can offer economic intuition to guide the interpretation of any results from the estimation of \eqref{eq:rf}. Third, even though an experimental design can confer a causal interpretation onto an elasticity, using an associated model to conduct counterfactuals can offer further causal inference along different dimensions, as the elasticity may have deeper underlying determinants that can be further understood using the estimation of an economic model \eqref{eq:st}. 

To illustrate these ideas, we turn to \cite{BriggsCesariniLindqvistOstling2021}, which tackles three related research questions. What happens to stock market participation after cash windfalls? Why are some households not participating in the stock market?  How large are the costs preventing them from doing so?

The reduced-form analysis relies on the institutional setting of Swedish lottery prizes, which have two interesting features. First, winning is itself a random event, and second, the form of payment---lump sum or annuity---is randomly assigned.  Thus, examining the effect of winning the lottery on stock market participation can distinguish the qualitative effect of a one-time entry cost into the market from a per-period participation cost.  The statistical models, \eqref{eq:rf}, include a regression of a dummy for stock market participation on different measures of lottery winnings, along with further specifications that employ interaction terms that indicate the type of prize payment.  
 
The results are interesting. A 150,000 US dollar windfall from lottery wealth increases the probability of stock ownership in post-lottery years by 4\%.   The effect is concentrated in previous stock market non-participants and in those winners who receive lump-sum prize payments. This last result shows that household stock market non-participation is influenced by a one-time entry cost instead of a per-period participation cost.

The structural part of the paper is an exercise in estimating the parameters of a life-cycle model with stock-market participation choice, participation start-up costs, and an unexpected lottery prize windfall. The decision rules from this environment can be characterized generally by \eqref{eq:st}.  The simulated minimum distance estimation uses the elasticities from the first part of the paper as a feature of the data that the model should strive to match. This integration is important because otherwise, the model counterfactuals would not be relevant for the well-identified reduced-form estimation that precedes them.

The results show that the entry cost for pre-lottery equity market nonparticipants that best fits the data is over 31,000 USD. However, even this cost cannot fully reconcile the small amount of participation. While a model with behavioral biases also fails in this regard, a model with downwardly distorted beliefs can reconcile about half of the difference between actual and model-generated participation. Interestingly, additional survey evidence backs up this model-based evidence. 

What do we learn from a combination of methods that we could not learn otherwise? The reduced form setting of random-sized lottery prizes provides an exogenous shock to household income that can identify the directional effect of wealth on participation, as well as the type of stock market participation cost. The model estimation then allows for the quantification of this cost, as well as the elimination of possible explanations for nonparticipation. These last exercises are also causal in nature, as they point to specific investor features as the driving forces behind their nonparticipation.

\section{Simplification}

The second type of integration of reduced-form and structural methods involves simplifying the model using a reduced-form regression to reduce complexity.  This strategy is useful for highly complex models if it is possible to simplify part of the model whose mechanism is too complicated to add to the current model and, importantly, does not affect the results of other parts of the model.  This type of strategy can itself be broken down into two subtypes, the first of which is widely used in industrial organization and dates back to the original notion of a reduced-form as being derived formally from an economic model.  One example is the demand estimation techniques based on the work of \cite{BerryLevinsohnPakes1995}, which are simply random-coefficient, instrumented logistic regressions that are formally derived from random utility models. Because of they are derived from a formal model, the parameter estimates can be used to conduct counterfactual experiments based on the model.  In the case of demand estimation, the statistical model \eqref{eq:rf} is the logit regression and the economic model \eqref{eq:st} is the random utility model. Examples in corporate finance and banking abound \citep[e.g.,][]{Xiao2020,DiMaggioEganFranzoni2021,EganLewellenSunderam2021}.   

The conditional choice probability (CCP) methods in \cite{HotzMiller1993} and the policy function approach in \cite{BajariBenkardLevin2007} are two further examples. While these methods have not been widely adopted in corporate finance, \cite{MatvosSeru2014} use the methods in \cite{BajariBenkardLevin2007}, and \cite{KangLoweryWardlaw2015} use CCP methods. 

This second paper clearly illustrates the advantages of the clever simplification that CCP methods confer. The research question is policy-relevant: how do regulators choose whether to close a troubled bank?    The paper constructs a dynamic discrete model of the choice to close a troubled bank or to leave it running. The tension in the model stems from a tradeoff between possible hikes in deposit insurance costs from delaying closure and the disutility regulators experience from closing banks.  The decision rule from this model then corresponds to \eqref{eq:st}. 

\cite{HotzMiller1993} show that in a broad class of models of this type, the difference in an agent's utility from each decision is proportional to the probability of that decision. These probabilities can be estimated via reduced-form logistic regressions, which correspond to \eqref{eq:rf}. These auxiliary regressions greatly simplify computation, as the whole dynamic program does not need to be solved repeatedly to estimate model parameters.  With these relative probabilities and thus relative utilities in hand, the model can be used to compute counterfactuals.  In the context of regulatory bank closure,\cite{KangLoweryWardlaw2015} find that political influence and a desire to defer costs matter most for the delayed closure of troubled banks. 
		
This type of use of auxiliary regressions to simplify parameter estimation rests on clever formal derivations. The second type of integration of reduced-form methods is informal.   For example, \cite{CooperHaltiwanger2006} use SMM to estimate the parameters of a single-agent dynamic investment model to ascertain whether fixed or convex adjustment costs are more important for explaining establishment-level investment. The investment decision rules from this model take the form of \eqref{eq:st}. However, they do not estimate all of the parameters using SMM to avoid lengthy computations. Instead, they estimate the production function parameters using an instrumental variables regression, which is encompassed by \eqref{eq:rf}. They then plug these parameters into the dynamic investment model and estimate the rest of the parameters using SMM. 

Another example is in \cite{WangWhitedWuXiao2022}, who measure the extent to which market power and regulatory frictions affect the pass-through of policy rates to bank lending decisions?  This question is counterfactual in nature and is difficult to address with reduced-form methods. Market power is hard to measure, and the introduction of regulatory frictions affects all financial institutions at once, so natural control groups do not exist.  

While reduced-form methods are unsuitable, to see how they can aid simplification, it is useful to outline the model.  Any theoretical structure whose decision rules look like \eqref{eq:st} and that is also useful for answering this question needs to be complicated, as it needs to embody dynamic optimization by banks, imperfect competition between banks, and equilibrium between borrowers, lenders, and banks. Each of these three model features requires solving for a fixed point, all of which are nested.  

However, equilibrium can be simplified via the estimation of loan and deposit elasticities using the methods in \cite{BerryLevinsohnPakes1995}.  Plugging these elasticities into the model implies that markets automatically clear, as interest rate choices by banks imply optimal quantities demanded from the estimated elasticities. 

This model simplification means that the parameters can be estimated with simulated minimum distance.  The model counterfactuals computed using these parameter estimates produce two main results. First, deposit market power matters a great deal for the pass-through of movements in the federal funds rates to bank customers, but so does bank capital regulation. Second,  bank-capital regulation and deposit market power interact to generate a reversal rate, below which cuts in the federal funds rates depress optimal bank lending.
	
\section{Internal Validity Extension} 	
	
The third type of integration consists of estimating the parameters of a model and conducting counterfactuals to extend the internal validity of the results from a reduced-form exercise. For example, a researcher might be interested in assessing the general equilibrium consequences of the estimates of elasticities or in predicting the effect of non-compliers in a quasi-experimental design. 

One example of this type of integration is in \cite{BernsteinColonnelliMalacrinoMcQuade2021}, who try to identify the characteristics of entrepreneurs that respond to local economic shocks. Using Brazilian matched employer-employee data, the paper exploits municipality variation in exposure to commodity price shocks to examine the response of new firms to these investment opportunities and the characteristics of the entrepreneurs who respond to these opportunities.  The identification strategy is to generate Bartik-style variation in the investment opportunities of municipalities using lagged industrial composition. The identifying assumption is that these initial sectoral differences in exposure are effectively randomly assigned, conditional on municipality fixed effects and other controls. The actual empirical model is of the form \eqref{eq:rf}, in which a municipality level outcome is regressed on the Bartik-style measure of investment opportunities, fixed effects, and controls. 

This setting offers two main results.  Young potential entrepreneurs respond to these shocks more than older ones, and this strong response is concentrated in municipalities with more developed banking sectors and a skilled population. While well-identified, this analysis is limited in scope because it cannot provide answers to questions such as the effects of changing the population composition. 

To answer questions of this sort, the paper estimates the parameters of a  dynamic discrete choice problem between wage employment and entrepreneurship. The decision to choose the type of employment can be characterized by \eqref{eq:st}.  In the model, selection into entrepreneurship depends on the dispersion in a  taste parameter that can be disciplined with the data.   Counterfactuals based on the model estimates show that a 10\% increase in the fraction of young people increases firm creation by around 2\%. 
Moreover, if this increase in the young is concentrated in the non-educated, this effect is muted by 30\%. 

These last two results could not have been obtained from the reduced-form regressions because estimation of these effects would require sufficient variation in the fraction of young people in a municipality. More generally, counterfactuals from models with estimated parameters are particularly informative when data are limited.  

Another example of this type of integration is in \cite{Antill2021}, which asks whether bankruptcy liquidation decisions are efficient and whether inefficient liquidation reduces creditor recovery. To instrument for the choice of the bankruptcy outcome, the paper uses the bankruptcy judge’s preference towards liquidations, measured as the out-of-sample fraction of bankruptcies the judge presided over that ended up in Chapter 7. Thus, the empirical model \eqref{eq:rf} is a simple instrumental variables regression.  The paper finds that for compliers who are close to the marginal threshold of liquidation versus emerging, the average liquidation reduces creditor recovery by 58 cents on the dollar. Of course, the limitation of this empirical design is that the result is a local average treatment effect that only applies to compliers, that is, those observations that are affected by the instrument. In the context of bankruptcy, the paper argues that compliers are likely to be contentious cases and thus not representative of the entire population of bankruptcies. 

To extend the inference from the compliers to the non-compliers, the paper uses a generalized \cite{Roy1951} model in which a binary choice (liquidation versus reorganization) depends only on the outcome of the choice.  For this type of model, the decision rule \eqref{eq:st} is just this simple choice.  In a way similar to the Heckman model, this framework allows for a sample-selection correction to extend the results to non-compliers. This exercise can answer how expected creditor recovery would change on average for all observations dcif form-of-exit decisions had been made by a statistical model that used past data on court filings. The result is that a statistical model could improve recovery by 12 cents on the dollar.

\section{Motivation and External Validity}

Most dynamic models provide a plethora of predictions regarding relationships between exogenous or endogenous variables in a data set. While some of the predictions can be used to estimate model parameters, especially in the context of indirect inference, usually many other predictions are not. These separate predictions can be used to motivate building a model in the first place or as a formal or informal check on external validity.  These ex-ante and ex-post checks share the feature that they lend credibility to any model counterfactuals. This type of research design has become the de facto blueprint in much macroeconomic research, and it also has many examples in corporate finance.

One example is \cite{PapanikolaouFrydman2018}, which starts with the observation that both executive pay and the gap between worker and executive pay have risen sharply over the last century.  Thus, the first reduced form model of the form of \eqref{eq:rf} are simple univariate trends. The paper then uses a dynamic equilibrium model of executive compensation and firm innovation to rationalize these trends. In this model, executives have two types of skills. They can manage existing projects, or they can find new projects. Both types of skills get compensated. In contrast, workers only have the first skill. Thus, when the value of new projects rises, the value of finding these projects also rises, and so does the compensation for that skill.  The decision rules from this model of the form \eqref{eq:st} are rules that dictate optimal executive and worker compensation and optimal investment in innovative technology as a function of the current firm state. 

SMD estimation of the model uses benchmarks related to compensation and to innovation but not to the connection between the two.  Thus, a sensible check of external validity is the model's prediction of a positive relationship between executive pay and growth opportunities. The paper finds cross-sectional evidence that executives are paid more in firms with more growth opportunities and that time-series fluctuations in aggregate executive pay are correlated with aggregate growth opportunities. The methodology for this external check is a simple OLS regression, which fits under the umbrella of \eqref{eq:rf}.  

Another example is in \cite{IvanovPettitWhited2021}, which tackles the old question of the effect of taxes on firm leverage choices. The first part of the paper uses a staggered difference-in-difference setting around changes in state corporate tax rates and data on small, private, U.S.\ firms.  This research design is implemented with a simple OLS model and thus is a special case of \eqref{eq:rf}. This part of the paper produces elasticities that have a causal, directional interpretation. Corporate leverage increases following tax cuts and decreases following tax hikes. 

This result is puzzling, as it contradicts the traditional intuition that the tax benefit of debt makes leverage more attractive when corporate income taxes rise, not less.  The structural part of the paper then serves to offer intuition for this result, and it is in this sense that the causal empirical results motivate the structural estimation. 

The model is a dynamic equilibrium model of an economy in which heterogeneous firms make factor demand decisions and are financed by internal profits, cash savings, and external risky debt. In the model interest expense on this debt is tax-deductible.  The model solution, a special case of \eqref{eq:st}, is a decision rule that dictates optimal leverage and factor demand decisions as a function of the firm's current state, as well as the corporate tax rate. The negative effect of taxes on leverage arises endogenously because taxes lower default thresholds, which then lead to higher risky interest rates on corporate debt. In the face of higher external financing costs, firms optimally borrow less. This effect can offset the tax benefit for those firms that have sufficiently high leverage so that default is possible in some states of the world. 

The paper then does two different external validity checks, both of which are implemented with OLS regressions. First, although the causal elasticity is not used as a benchmark to estimate the model parameters, the model generates an elasticity with the same sign and a slightly smaller magnitude.  Second, as auxiliary evidence for the default mechanism, the paper finds a significant decrease in collateral requirements following tax cuts, that is, loan terms become more generous.

\section{Sample Selection}

The final method for integrating reduced-form and structural empirical consists of using the estimation of an economic model to treat the problem of sample selection in a regression. This type of work has a long history in economics.  For example, a Heckman correction is a regression, as in \eqref{eq:rf}, paired up with a probit, which is itself an outcome of a random utility problem. In this simple economic model, which is a special case of \eqref{eq:st}, the agent chooses to stay in the sample if their utility exceeds a threshold.  The Heckman model is simple because a normality assumption means that one can use a control-function approach instead of joint estimation of both the probit and the regression. However, the selection can be much more elaborate and thus economically interesting. 

An elegant example is \cite{Sorensen2007}, who asks why start-up companies funded by more experienced venture capitalists (VCs) are more likely to go public. Unfortunately, a simple regression of the going-public decision on venture capital experience, as in \eqref{eq:rf} is unlikely to answer the question because firm characteristics that cause them to be matched with a VC are in the error term. This question is thus a classic example of treatment versus selection.  This problem could be addressed via simple regression techniques if all of the variables that influence the sorting of companies to VCs were observable, but scanty data on private start-up firms make this problem real.

\cite{Sorensen2007} solves this problem not by using a simple random utility model as the underlying theory for the structural component of the estimation, but a two-sided matching model. Each VC can have more than one match, but each company can have only one VC. The matches a VC gets depend on competition from other VCs, so if a new VC enters the market, the existing VCs will have to settle for worse companies. The equilibrium concept is stability: perturbing the matching outcome would make any company's valuation worse. And the solution is a matching rule, as in \eqref{eq:st} that assigns firms to VCs.   

With appropriate distributional assumptions, the matching rule can be used to derive a likelihood function that can then be used for joint estimation of the matching rule and the reduced-form regression of the going public decision on VC experience. The paper uses Markov Chain Monte Carlo methods to maximize the likelihood function and estimate parameters. 

The simple reduced-form probit results show that an investment by the more experienced investor is 82\% more likely to succeed than an investment by an inexperienced investor. However, counterfactuals from the structural estimation show that only about one-third of this effect is due to experience. The rest is due to sorting.  This second result could not have been obtained by reduced-form methods alone because of data limitations, so this example further illustrates the usefulness of structural estimation when data are insufficient for answering a question.

\section{Conclusion} 

Empirical corporate finance, banking, and personal finance have largely been divided into two groups along methodological lines: reduced-form and structural. Many examples of interesting and well-executed studies exist in both camps. For example, \cite{BennedsenNielsenPerezGonzalezWolfenzon2007} offers a clean instrument to answer a question about family firms, and  \cite{SchrothSuarezTaylor2014} use estimation of a global games model of bank runs to understand the importance of various policies to prevent bank runs.  While there has been substantial disagreement, seen more often in referee reports than in published pieces, about the suitability of these two broad classes of methods, this debate is unfounded, as both methods can offer useful insights on their own. This paper has also sought to argue via example that these two types of methods can be complementary. 

The examples are varied. Estimation of the parameters of an economic model can offer insight into the underlying forces that determine the sign and magnitude of reduced-form slope coefficients. Similarly, estimation of model parameters can extend the external validity of reduced form results or solve sample selection problems. Parenthetically, the cross-sectional tests seen in many reduced-form papers to uncover underlying economic mechanisms are a form of structural estimation, albeit one in which the model is a verbal model.  Conversely, reduced-form estimation can offer external validity for models.  Undoubtedly, future researchers will find creative ways to mix and match these two broad classes of methods.


\newpage

\clearpage
\singlespacing


\end{document}